\documentclass[singlecolumn, preprintnumbers,amsmath,amssymb,balancelastpage]{revtex4}
\pdfoutput=1

\usepackage{graphicx}
\usepackage{dsfont}
\usepackage{amsmath}
\usepackage{array}
\usepackage{epstopdf}
\usepackage{enumerate}
\usepackage{youngtab}
\usepackage{tensor}
\usepackage{braket}
\usepackage[normalem]{ulem}
\usepackage{slashed}
\usepackage[aligntableaux=center]{ytableau}
\usepackage[utf8]{inputenc}
\usepackage[
      colorlinks=true,
      linkcolor=blue,
      urlcolor=blue,
      filecolor=black,
      citecolor=red,
      ]{hyperref}
\usepackage{braket}
\usepackage{mathtools}
\usepackage{rotating}
\usepackage{tabularx}
\newcolumntype{Y}{>{\centering\arraybackslash}X}
\newcolumntype{C}[1]{>{\centering\arraybackslash}p{#1}}
\usepackage{todonotes}
\usepackage{xcolor}
\definecolor{LightCyan}{rgb}{0.7,1,1}
\definecolor{Gray}{gray}{0.9}
\usepackage{xspace}

\newcommand {\be} {\begin {equation}}
\newcommand {\ee} {\end {equation}}

\newcommand {\bes} {\begin {equation*}}
\newcommand {\ees} {\end {equation*}}

\newcommand{\cN}{{\mathcal N}}
\newcommand{\cO}{{\mathcal O}}

\newcommand{\cM}{{\mathcal M}}
\newcommand{\cT}{{\mathcal T}}

\newcommand{\beq}{\begin{equation}}
\newcommand{\eeq}{\end{equation}}

\def\ie{\begin{equation}\begin{aligned}}
\def\fe{\end{aligned}\end{equation}}

\def\<{\langle}
\def\>{\rangle}

\DeclareMathOperator{\sgn}{sgn}

\def\beg{\begin{equation}\begin{gathered}}
\def\eeg{\end{gathered}\end{equation}}

\def\bea{\begin{equation}\begin{aligned}}
\def\eea{\end{aligned}\end{equation}}

\newcommand{\cc}{\mbox{c.c.}}
\newcommand{\mm}{M}
\newcommand{\blu}[1]{{\color{black}#1}}

\begin{document} 

\author{William Witczak-Krempa}
\affiliation{D\'epartement de physique, Universit\'e de Montr\'eal, Montr\'eal (QC), H3C 3J7, Canada}
\affiliation{Centre de Recherches Math\'ematiques, Universit\'e de Montr\'eal; Montr\'eal (QC), H3C 3J7, Canada}
\affiliation{Institut Courtois, Universit\'e de Montr\'eal, Montr\'eal (QC), H2V 0B3, Canada}

\title{Dirac quantum criticality and the web of dualities}

\begin{abstract}
We grow the web of dualities for conformal field theories (CFTs) in 2+1 spacetime dimensions to include quantum critical transitions of Dirac fermions. Our construction uses the seed duality, equating a free Dirac fermion with a complex boson coupled to a Chern-Simons gauge field, to express various Gross-Neveu-Yukawa (GNY) critical points of $N$ fermions in terms of the same number of complex bosons and partner gauge fields. 
In some instances, monopole operators appear explicitly in the dual action, such as for the superconducting transition of a single Dirac fermion, which is a CFT with emergent supersymmetry. We match phase diagrams, symmetries and their anomalies. By further gauging certain symmetries, we derive offspring dualities. For example, starting with the duality for the regular GNY transition of 2 Dirac fermions, we arrive at the duality between a N\'eel-to-VBS deconfined quantum critical point and a gauged-Gross-Neveu transition. Finally, we use a large-$N$ expansion for monopole operators, and compare with known scaling dimensions in the simplest single-flavour GNY transition.    
\end{abstract}

\maketitle
\tableofcontents
\nopagebreak

\section{Introduction}
The discovery of dualities between members of an important family of quantum critical systems, conformal field theories (CFTs) in 2+1 spacetime dimensions, has generated numerous new insights into quantum criticality, topological phases, quantum spin liquids, and anomalies~\cite{Polyakov88,maissam,Aharony2015,Seiberg2016,Karch:2016sxi,karch-more,Metlitski2017,Wang2017,review,Cenke2017,Benvenuti2018}.
In particular, the conjectured seed duality~\cite{Seiberg2016,Karch:2016sxi} between a single Dirac fermion and a gauged Wilson-Fisher boson with Chern-Simons (CS) coupling generates a rich web of dualities, that includes the celebrated particle-vortex duality of the superfluid-insulator transition~\cite{review}. A central role in the dualities is played by monopole (instanton) operators of U(1) gauge fields. These are tunneling events of a quantum of magnetic flux of an emergent (compact) gauge field. In the seed duality, the Dirac fermion operator is dual to the gauge-invariant monopole, which is the bound state of a flux instanton and the boson. In the particle-vortex duality, the regular Wilson-Fisher boson becomes dual to the monopole in the abelian Higgs dual. Results for monopoles also gave  surprising evidence for the web of dualities: large-$N$ calculations for monopole scaling dimensions extrapolated to $N=1$ gave excellent agreement with the seed and particle-vortex dualities~\cite{Dyer15,Chester22}. 

However, relatively little attention has been given to quantum critical transitions involving Dirac fermions~\cite{karch-more,Metlitski2017,review,xu23}. These transitions play a central role in the study of quantum phases of matter, ranging from graphene to topological materials~\cite{joseph21}. 
For instance, a duality for a Gross-Neveu (GN) transition for a single Dirac cone was obtained~\cite{karch-more} from the seed. Below we review the construction, and present a slightly different formulation via a UV-complete Yukawa interaction. This allows us to match the phase diagram and symmetries. We then apply this procedure to the superconducting transition of a single Dirac cone~\cite{Balents98,Strassler03,Lee06,Grover13,Bobev15,wk15}, the supersymmetric Wess-Zumino CFT; monopoles play a key role in the duality. We then study various Gross-Neveu-Yukawa (GNY) transitions with $N=2$ Dirac cones, by using 2 copies of the seed duality and an appropriate interaction term. This includes O(4) symmetric, chiral Ising, chiral XY and chiral Heisenberg GNY CFTs. The dual bosonic QED theory has 2 dynamical gauge fields, and an appropriate interaction term. We then generalise to $N$ Dirac fermions, and perform a check at $N\to\infty$. We further investigate new dualities that follow from gauging the global symmetries. In particular, we arrive at the duality~\cite{Wang2017} between a N\'eel-to-valence bond solid (VBS) deconfined quantum critical point~\cite{dqcp}, and a gauged-Gross-Neveu transition (fermion QED-Gross-Neveu theory). Finally, we use a large-$N$ expansion for monopole operators, and compare with known scaling dimensions in the simplest single-flavour GNY transition.

\section{Quantum criticality of a single Dirac fermion} \label{sec:1dirac}
\subsection{Seed duality}
The seed duality in a recently discovered web of dualities~\cite{Seiberg2016,Karch:2016sxi} between Conformal Field Theories (CFTs) connects two seemingly different theories: a free Dirac fermion and a complex boson with self-interactions coupled to a gauge field with Chern-Simons (CS) interaction:
\begin{align} \label{seed}
    L_f &= \bar\psi i\slashed{D}_{\!A} \psi \\
    L_b &= |D_a z|^2 - |z|^4 + \frac{1}{4\pi}ada   + \frac{1}{2\pi}A d a \nonumber
\end{align}
where the fermion is a single 2-component Dirac spinor coupled to a background (spin$_c$) gauge field $A$; it could be realized at the boundary of a 3+1D topological insulator protected by time-reversal and charge conservation. The gauge covariant derivative is $D_{\!A} = \partial - i A$, and $\slashed D=\gamma^\mu D_\mu$ with 
2-by-2 Dirac $\gamma$-matrices. The Chern-Simons (CS) term is $\tfrac{k}{4\pi} ada= \tfrac{k}{4\pi}\epsilon^{\mu\nu\rho} a_\mu\partial_\nu a_\rho$, where $k\in\mathbb Z$ is the level. Its role is crucial in the above boson-fermion duality as it attaches $k=1$ unit of flux to the complex boson $z$, thus converting it to a fermion. 

\subsection{Gross-Neveu-Yukawa transition}\label{sec:gny1}
We begin with the simplest quantum critical theory of a single 2-component Dirac fermion: the GNY theory with global O(2) symmetry. Conformal bootstrap gave rigorous results~\cite{archi} for this CFT, to which we shall come back below.
Using the seed \eqref{seed}, the bosonic dual was obtained in \cite{karch-more}, but we shall use a slightly different formulation that we find more convenient for the analysis of the neighboring phases. We begin with the seed \eqref{seed}, 
and use the correspondence between the mass terms
\begin{align} \label{seed-mass}
    \bar\psi\psi \leftrightarrow |z|^2
\end{align}
to add a non-dynamical source $\phi$ (it is a spacetime dependent field that is not integrated over in the path integral) for these operators on both sides:
\begin{align} \label{gn}
    L_f' &= L_f - h\phi\bar\psi\psi -r \phi^2 - \phi^4  \\
    L_b' &= L_b - h\phi z^\dag z -r \phi^2 - \phi^4  \nonumber
\end{align}
We have added additional terms for the source that will be useful in the next step. We shall leave the standard kinetic term $(\partial\phi)^2$ implicit. We now make $\phi$ dynamical. One recognizes $L_f'$ as the GNY transition obtained by tuning $r$ to zero. The order parameter field $\phi$ is a pseudoscalar, whose condensation when $r<0$ leads to a mass for the fermion. $L_b'$ is the ``bosonized'' dual of this transition, which is a critical point since only the parameter $r$ needs to be tuned to reach criticality. Indeed, $|z|^2$ is forbidden being odd under time-reversal; not taking this fact into account would lead one to call $L_b'$ tri-critical. As in the seed duality, the fermion is dual to the gauge invariant monopole of minimal magnetic charge (flux):
\begin{align} \label{m1}
    \psi \leftrightarrow z^\dag \mm_a 
\end{align}
where $\mm_a$ inserts a $2\pi$ flux of $a_\mu$. It is dressed by a boson creation operator due to the CS term. In other words, the continuous U(1)
symmetry on the fermion side is realized as the topological U$(1)_T$ symmetry on the boson side, under which the monopoles are charged.  

In \cite{karch-more}, a slightly different formulation for the dual of $L_f'$ at criticality was obtained. The idea is to first decouple the $|z|^4$ in $L_b$ with a Hubbard-Stratonovich field to $\sigma |z|^2$.
Then $|z|^2$ in the duality between mass operators \eqref{seed-mass} can be replaced with $\sigma$. Finally, adding a term $\phi\sigma$ (this is the only $\phi$ dependence) allows one to shift $\phi$ to eliminate $\sigma|z|^2$, and integrating out $\phi$ gives 
\begin{align}
    |D_a z|^2+\frac{1}{4\pi}ada+\cdots
\end{align}
where the ellipsis represents higher order potential terms for $z$ unspecified in \cite{karch-more}. 
We shall find it more convenient to keep the full Yukawa form, as we now discuss.

\emph{Phase diagram}---When $r>0$, $\phi$ is gapped on both sides, and can be neglected since $\phi^2$ is relevant under RG. We are thus left with a free massless Dirac fermion in the IR, and its dual version \eqref{seed}. When $r<0$, $\langle\phi\rangle=\phi_0$ spontaneously breaks time-reversal which gives the fermion in $L_f'$ a mass term $\phi\bar\psi\psi\to \phi_0\bar\psi\psi$. There are two degenerate vacua: $\phi_0>0$ and $\phi_0<0$. 
On the boson side, the $\phi$ condensate gives the mass term $\phi |z|^2\to \phi_0|z|^2$. This matches the fermion side owing to the identification of mass operators \eqref{seed-mass}. Explicitly, for $\phi_0>0$, $z$ becomes gapped, and the effective Lagrangian is $-\tfrac{1}{4\pi}AdA-2\mbox{CS}_g$. For the other vacuum, $\phi_0<0$, the negative mass term for $z$ forces it to condense $\langle z\rangle\neq 0$, which Higgses out $a_\mu$, leaving a fully IR trivial theory without CS terms in its effective response Lagrangian. These two degenerate vacua exactly match the fermion theory.  
We can also consider deforming by a mass term $L_f'\to L_f' - m\bar\psi\psi$ in the symmetric phase. Since $\phi$ is gapped and $\langle\phi\rangle=0$, it can be ignored in the IR. One gets a massive fermion with mass $m$, which can be positive or negative. On the boson side, $L_b'\to L_b'-m|z|^2$. 
We can again ignore $\phi$ in the IR, so we fall back on the seed duality, and get exact agreement. 

\emph{Symmetries}---As was discussed in \cite{Seiberg2016,Karch:2016sxi}, all symmetries of the seed duality match. Since the $\phi$ is a pseudoscalar, the $\phi$-terms that we added in \eqref{gn} respect all symmetries, and thus $L_f'$ and $L_b'$ have the same symmetries. Let us review how this works for time-reversal $\cT$. A single Dirac fermion is $\cT$-invariant up to an anomaly, thus the $N=1$ GNY theory shares this anomaly:
\begin{align} \label{trs}
    L_f' \xrightarrow{\cT} L_f' + \frac{1}{4\pi}AdA +2\mbox{CS}_g
\end{align}
\blu{where $2\mbox{CS}_g$ is a gravitational CS term that encodes the thermal Hall conductivity~\cite{Seiberg2016}. }
On the boson side, time-reversal acts as a particle-vortex duality transformation: $\cT(A)=-A$, $\cT(a)=-\hat a$, $\cT(z)=\hat z$, and $\cT(|z|^2)=-|\hat z|^2$. We note that the action is indicated for 
the spatial components of the gauge fields, whereas the time-components $a_0, A_0$ transform with the opposite sign. 
Further, $\phi$ being a pseudoscalar, $\cT(\phi)=-\phi$. This shows that all $\phi$-terms are invariant. Time-reversal changes the sign of both CS terms in $L_b$ \eqref{seed}. However, $\cT(L_b)$ is dual via particle-vortex duality to $L_b$ up to the same anomaly as in \eqref{trs}. We thus obtain that both $L_f'$ and $L_b'$ share the same time-reversal symmetry (up to an identical anomaly). 

\subsection{\blu{Superconducting transition}} 
A Dirac fermion possesses another important instability besides spontaneous mass generation: pairing. The charge-2  ``Cooper pair'' operator $\psi^2\equiv \psi_\alpha \epsilon^{\alpha\beta}\psi_\beta$, where we have explicitly written out the spinor indices, is a Lorentz scalar. $\epsilon^{\alpha\beta}$ is the anti-symmetric matrix $\epsilon^{12}=-\epsilon^{21}=1$. The condensation of $\psi^2$ describes an s-wave superconducting (or more precisely, paired superfluid) state of fermions. Its dual is the flux $4\pi$ gauge-invariant monopole operator
\begin{align} \label{m2}
    \psi^2 \leftrightarrow (z^\dag)^2 \mm_a^2\equiv \cM_2 
\end{align}
We have attached two $z^\dag$ operators since there is an insertion of two quanta of flux. We now follow the same procedure to obtain a dual formulation of the quantum critical point between the Dirac semimetal and the superconductor. We first add a source term $\phi \psi^2+\mbox{c.c.}$, where $\phi$ is a complex scalar background charge-2 field that transforms oppositely to $\psi^2$ under $A$ gauge transformations. We do the same on the boson side, and add pure-$\phi$ terms as above to get:
\begin{align} \label{wz}
    L_f' &= L_f - (h\phi\psi^2+\cc) -r |\phi|^2 - h^2|\phi|^4 + |\partial\phi|^2 \\
    L_b' &= L_b - (h\phi \cM_2+\cc) -r |\phi|^2 - h^2|\phi|^4 + |\partial\phi|^2   \nonumber
\end{align}
\blu{This time, we explicitly wrote the $|\partial\phi|^2$ term. With $\phi$ non-dynamical, this is a deformation of the seed duality \eqref{seed} that probes how the free fermion, and its dual, respond to a superconducting order imposed by hand; we could take $\phi$ constant in that case.}
\blu{In order to investigate a new critical transition,} we now promote $\phi$ to a dynamical field. On the fermion side, we recognize $L_f'$ as the quantum critical transition between a Dirac semimetal and superconductor obtained by tuning $r$ to zero. 
This so-called Wess-Zumino CFT has an emergent $\cN=2$ spacetime SUSY that rotates the fermion into the boson $\phi$, as reviewed in \cite{Strassler03}. The couplings of the two interaction terms involving $\phi$ in $L_f'$ are simply related at the \blu{new IR} fixed point.  
\blu{We note that when $h=r=0$, by construction we have a non-interacting supersymmetric theory of a non-interacting fermion and a complex scalar on both sides of the duality.} 

The bosonic gauge theory now explicitly involves the gauge invariant $4\pi$-flux monopole operator $\cM_2$. \blu{Since the U(1) symmetry on the boson side is the topological flux-conservation symmetry, we interpret the new dynamical field $\phi$ as new monopole operator with U(1)$_T$ charge $-2$. We could thus write it as 
\begin{align}
    \phi= (\mm^\dag_a)^2 w
\end{align} 
where $\mm^\dag_a$ inserts a flux of $-2\pi$, and $w$
is a new complex scalar that has gauge charge $-2$. The interaction term $\phi\cM_2$ is in the zero-flux sector, so we could write it as $w\,(z^\dag)^2$ at low energy. Since $w$ is a new degree of freedom, the possibility that $\phi$ acquires a scaling dimension lesser than that of $\cM_{\pm 2}$ (which is built with two $z$ bosons) is not unreasonable, and required by SUSY. }

\emph{Phase diagram}---When $r>0$, $\phi$ is uncondensed and gapped on both sides, and can be ignored. We thus get the seed duality for a massless Dirac fermion. 
For $r<0$, a Cooper pair condensate appears $\langle\phi\rangle\neq 0$, which spontaneously breaks the U(1) charge-conservation symmetry down to $\mathbb Z_2$.
Indeed, the pair condensate is invariant under $\psi\to -\psi$. A Goldstone mode is generated, while the fermion becomes gapped. On the bosonic side, the $\phi$-condensate gives a Goldstone mode; it also sources the monopole operator $\cM_2$, which acquires an expectation value. This breaks the U(1)$_T$ symmetry down to 
$\mathbb Z_2$, nicely matching what occurs on the fermion side. The monopole condensation leads to a Higgs effect, gapping out both $z$ and $a$, which leaves a single Goldstone mode in the IR. The SUSY rotates the minimal monopole $\cM_1=z^\dag \mm_a$ into $\phi$. 

\emph{Symmetries}---As for the GNY transition, the usual (non-SUSY) symmetries can be seen to match from the fact that the added $\phi$-terms in \eqref{wz} do not break the symmetries of the unperturbed Lagrangians. The SUSY is emergent, and rotates $\cM_1$ into $\phi$ as was discussed above. 

\section{Quantum criticality of two Dirac fermions} \label{sec:2dirac}
Let us now see how the ideas above can be adapted to quantum critical points of $N=2$ Dirac fermions.

\subsection{Ising Gross-Neveu} \label{sec:ign}
Let us begin with the ``chiral'' Ising GNY transition (IGN) with $\mbox O(2)^2\!\rtimes\!\mathbb Z_2$ global symmetry, as reviewed in \cite{archi}. In its fermionic definition, it is realized by coupling a real scalar order parameter field to the staggered mass operator: $\bar\psi_1\psi_1-\bar\psi_2\psi_2$. Using the seed duality for each fermion, the dual of the staggered mass operator becomes $|z_1|^2-|z_2|^2$. We can apply the procedure described above by coupling the staggered mass to a background scalar $\phi$, add a background $\phi$-potential, and make $\phi$ dynamical:
\begin{align} \label{ign}
    L_f' &= L_f[\psi_1] + L_f[\psi_2] - h\phi(\bar\psi_1\psi_1-\bar\psi_2\psi_2) -r \phi^2 - \phi^4  \\
    L_b' &= L_b[z_1,a_1] + L_b[z_2,a_2] - h\phi( |z_1|^2-|z_2|^2) - 
    r \phi^2 - \phi^4  \nonumber
\end{align}
We emphasize that there are now two gauge fields in $L_b'$. 
We left implicit the coupling to a background gauge field $A_i$ in $L_f[\psi_i]$; same for $L_b[z_i,a_i]$.  Both sides possess a global $\mathbb Z_2$ symmetry that interchanges $1\leftrightarrow 2$ and sends $\phi\to -\phi$. The two fermions are dual to the gauge invariant monopoles labelled by $i=1,2$:
\begin{align} \label{psi2}
    \psi_i \leftrightarrow z^\dag_i \mm_{a_i}\equiv \cM_1^{(i)} 
\end{align}

\emph{Phase diagram}---The $\mathbb Z_2$ symmetry exchanging $1\leftrightarrow 2$ is spontaneously broken when $r<0$. Indeed, the real scalar spontaneously breaks the $\phi\to-\phi$ symmetry, and gives masses of opposite signs to the two fermions. This leads to a fully gapped phase without  Goldstone modes. On the boson side, the $\phi$ condensate $\langle\phi\rangle=\phi_0$ gives masses of opposite signs to $z_1$ and $z_2$. The fluctuations of the $\phi$ are gapped, and can be ignored at low energy. We are left with the bosonic dual of two fermions with staggered masses, which matches $L_f'$. 
We can now examine the effective response in the broken symmetry phase. Defining new background gauge fields via $A_1=A+B$ and $A_2=A-B$, where $A$ is a spin$_c$ connection, the effective response on the fermion side is 
\begin{align} \label{ign-response}
   L^{\rm eff}= -\frac{1}{4\pi}(A\pm B)d(A\pm B) -2\mbox{CS}_g
\end{align}
where $\pm=\sgn \phi_0$. This leads to the following difference in the effective Lagrangian between the two vacuua:
\begin{align} \label{qsh}
    \Delta L^{\rm eff} = \frac{1}{\pi} AdB
\end{align}
which is a topological quantum spin Hall response~\cite{Cho2010}, where a ``charge'' current associated with $A$ generates a transverse ``spin'' current associated with $B$. On the boson side, only the $z_i$ boson with positive mass contributes to the effective response, and we get the same response as on the fermion side \eqref{ign-response}-\eqref{qsh}. This discussion will further apply to the disordered phase $r>0$ perturbed by a staggered mass term. It confirms the expectation that two massless Dirac fermions sit at the transition between an insulator with quantum spin Hall response, and a trivial one. 

In the disordered phase $r>0$, we can perturb with a symmetric mass term $L'_f-m(\bar\psi_1\psi_1+\bar\psi_2\psi_2)$. Since $\phi$ is gapped, we can ignore it low energy and set $h=0$. We thus have two free fermions with identical masses.
In the boson language, the perturbed Lagrangian is $L_b'-m(|z_1|^2+|z_2|^2)$. Since $\phi$ is gapped, we can again ignore it in the IR, giving the dual of two free fermions with equal masses.

\emph{Symmetries}---The continuous U(1)$^2$ symmetry in the IGN theory is realized via the topological symmetries of $a_1$ and $a_2$, $\mbox U(1)_T^2$. Besides the exchange $\mathbb Z_2$ symmetry already discussed above, each flavour has a $\mathbb Z_2$ charge conjugation symmetry that leaves all the $\phi$-terms in \eqref{ign} invariant. Regarding time-reversal $\cT$, different implementations exist. We take the usual time-reversal transformation for each flavour as given in the seed duality, but we also include the standard transformation on flavour ``spin'' space under which the staggered mass is odd (in flavour space, $\sigma_z\to-\sigma_z$). The operator $\phi$ is $\cT$-even, making the interaction term $\phi(|z_1|^2-|z_2|^2)$ time-reversal invariant.
We see that both sides are invariant up to the same anomaly. Alternatively, we could define $\cT$ without the flavour transformation, in which case $\phi$ would be odd. 

\subsection{Gross-Neveu-Yukawa transition with O(4) symmetry}
We now study the generalisation of the GNY transition studied in \ref{sec:gny1} to $N=2$ flavours. Adapting the procedure above, we get
\begin{align} \label{gn2}
    L_f' &= L_f[\psi_1] + L_f[\psi_2] - h\phi\bar\Psi\Psi -r \phi^2 - \phi^4  \\
    L_b' &= L_b[z_1,a_1] + L_b[z_2,a_2] - h\phi Z^\dag Z -r \phi^2 - \phi^4  \nonumber
\end{align}
where we defined 
\begin{align}
    \Psi=\begin{pmatrix}
\psi_1 \\ \psi_2 
\end{pmatrix}\,, 
\qquad 
   Z =\begin{pmatrix}
z_1 \\ z_2 
\end{pmatrix}
\end{align}
Eq.~\eqref{gn2} follows from the identification of the mass operators $\bar\psi_1\psi_1\leftrightarrow |z_1|^2$ given in \eqref{seed-mass}; the same holds for flavour $i=2$. 
The real order parameter field $\phi$ is a pseudoscalar (odd under $\cT$).

\emph{Phase diagram}---When $r>0$, $\phi$ becomes gapped 
and can be ignored in the IR: we obtain two copies of the seed duality, as expected for two massless Dirac fermions. 
When $r<0$, $\phi$ condenses giving both fermions identical masses.
On the boson side, the condensate $\langle\phi\rangle=\phi_0$ gives identical masses to the bosons. When $\phi_0>0$, the $z_i$ bosons become gapped, and we again find agreement owing to the seed duality, and the correspondence between the fundamental mass operators \eqref{seed-mass}. 
For the other choice of vacuum, $\phi_0<0$, the $z_i$ bosons have identical negative masses, which leads to their condensation, $\langle Z\rangle\neq 0$. Owing to the potential for $Z$, we get symmetric condensates $\langle z_1\rangle=\langle z_2\rangle$ that Higgs both $a_1$ and $a_2$, without additional Goldstone modes. We thus obtain the bosonic dual of two free fermions with identical negative masses. 

\emph{Symmetries}---The GNY CFT has a global O(4) symmetry. On the bosonic side, $L_b[1]+L_b[2]$ also has O(4) symmetry by virtue of being dual to two massless Dirac fermions, although it is not manifest.
The interaction term $\phi Z^\dag Z$ is also invariant under O(4) since both $\phi$ and $Z^\dag Z\leftrightarrow\bar\Psi\Psi$ are O(4) singlets, which shows that global symmetries match on both sides. $L_b'$ possesses time-reversal (up to an anomaly) since $Z^\dag Z$ is odd under $\cT$. 

\subsection{Heisenberg Gross-Neveu transition}
We now examine a cousin of the IGN theory studied above: the so-called chiral Heisenberg Gross-Neveu (HGN) transition that involves a 3-component scalar order parameter $\vec\phi=(\phi_x,\phi_y,\phi_z)$ that couples to $\bar\Psi\vec\sigma\Psi$, where $\sigma_l$ are the 3 Pauli matrices acting on flavour space (we could call it ``spin''). The IGN would obtain by setting $\phi_x=\phi_y=0$. The discussion below can be readily adapted to the chiral XY GNY transition by setting $\phi_z=0$.
We propose  the following duality for HGN
\begin{align} \label{hgn}
    L_f' &= L_f[\psi_1] + L_f[\psi_2] - h\vec\phi\cdot\bar\Psi\vec\sigma\Psi -r \phi^2 - (\phi^2)^2  \\
    L_b' &= L_b[z_1,a_1] + L_b[z_2,a_2] - h\vec\phi\cdot \bar\cM\vec\sigma \cM -r \phi^2 - (\phi^2)^2  \nonumber
\end{align}
where $\phi^2=\vec\phi\cdot\vec\phi$. The bilinear $\bar\Psi\vec\sigma\Psi$ would be dual to $\bar\cM\vec\sigma \cM$, where we defined the gauge invariant monopole doublet 
\begin{align}
    \cM =   \begin{pmatrix}
\cM_1^{(1)} \\ \cM_1^{(2)} 
\end{pmatrix}
\end{align}
$\bar\cM$ is the hermitian conjugate of $\cM$ times the appropriate matrix.
The need for invoking monopoles arises since $\bar\Psi\sigma_{x,y}\Psi$ involve the mixed mass term $\bar\psi_1\psi_2$, whereas $z_1^\dag z_2$ is not gauge invariant by itself. To check this operator identification, we will match the phase diagram and symmetries of both theories.  

\emph{Phase diagram}---For $r>0$, $\vec\phi$ is gapped, and disappears from the IR so that we recover 2 massless Dirac fermions on both sides of the duality. For $r<0$, $\vec\phi$ acquires an expectation value $\langle\vec\phi\rangle\neq 0$, which spontaneously breaks the SO(3) rotation symmetry and generates 2 Goldstone modes.
This occurs on both sides. Indeed, without loss of generality (invoking flavour rotation symmetry) let us assume that that the condensate is aligned along the $\sigma_z$ axis. On the boson side we get a mass term proportional to $\phi_0 (|\cM_1^{(1)}|^2- |\cM_1^{(2)}|^2)$. Now, the first term involves the product of a flux creation operator and and its corresponding annihilation operator for $a_1$. This product is in the zero-flux sector, and can be taken to be the identify at low energy; the same holds for the second term for flavour $i=2$. We can thus write the mass term as $\phi_0(|z_1|^2-|z_2|^2)$, which we recognize as the staggered mass of the IGN theory. As described in section~\ref{sec:ign}, this staggered mass gaps out one flavour and Higgses out the other, leading to the correct dual of two fermions with staggered masses. 

In the Dirac semi-metal phase $r>0$, we can perturb by a symmetric mass term $L_f'-m\bar\Psi\Psi$, which leads to two massive Dirac fermions with identical masses. Indeed, the $\phi$ field is gapped and can be ignored in the IR. On the boson side, this becomes $L_b'-m Z^\dag Z$. Since $\phi$ can be ignored, we recover the correct dual of two Dirac fermions with mass $m$. 

\emph{Symmetries}---The continuous symmetry group of HGN is the $\mbox{SU}(2)\!\times\!\mbox U(1)$ subgroup of SO(4) due to the interaction term.
In the bosonic gauge theory, the diagonal topological symmetry U(1)$_T$ under which both monopoles transform with the same phase,
\begin{align}
    \cM \to \cM e^{i\theta}\,,
\end{align}
realizes the U(1) of the fermion side. This transformation clearly leaves invariant the interaction term in $L_b'$. The SU(2) acts on the gauge invariant monopoles as it does on the fermions owing to the identification $\cM\leftrightarrow\Psi$, making the interaction term SU(2) symmetric.
Since $L_b[1]+L_b[2]$ has $O(4)$ symmetry, the full theory $L_b'$ has a continuous SU(2) symmetry. 
Regarding time-reversal, on the fermion side we can define it with the usual action on each flavour plus the action on flavour space such that $\vec\sigma\to-\vec\sigma$. 
In this case, the $\vec\phi$ field is time-reversal even, which also holds on the boson side. By construction the interaction term in $L_b'$ is even. Combining this with the time-reversal transformation of $L_b[1]+L_b[2]$, we get that both theories are $\cT$ invariant up to the same anomaly. 

\section{Many Dirac fermions}
We now generalise the procedure to quantum critical points of $N$ Dirac fermions. As an example, let us examine the duality for the GNY theory with O($2N$) symmetry:
\begin{align}
    L_f' &= \sum_{i=1}^N\left( L_f[\psi_i]  - h\phi\bar\psi_i\psi_i\right) -r \phi^2 - \phi^4  \\
    L_b' &= \sum_{i=1}^N\left( L_b[z_i,a_i] - h\phi z_i^\dag z_i\right) -r \phi^2 - \phi^4  \nonumber
\end{align}
The disordered phase at $r>0$ naturally gives $N$ massless Dirac fermions in the IR, and its seed-dual formulation. In $L_f'$, $r<0$ leads to a spontaneous mass generation for all the fermions $\phi_0\bar\psi_i\psi$, where the two vaccua correspond to the two signs for the $\phi$-expection value $\phi_0$. The same happens on the boson side, and no Goldstone modes are generated owing to the potential for $z_i$. One can further match the symmetries as we did for $N=2$. 

We shall now study the limit $N\to\infty$, in which the GNY CFT becomes exactly solvable. At large-$N$, one can ignore the $\phi^4$ term and work with the Yukawa interaction where we set $h=1$. On the boson side, the same holds. Working with the partition function of $L_b'$, we can first perform the path integral over $z_i$ and $a_i$, which decouple between different $i$. For a given $i$, $L_b[z_i,a_i]-\phi|z_i|^2$ is dual to a fermion coupled to a (spacetime-dependent) mass $\phi$. So the effective action for $\phi$ is $N S_1[\phi]$, where $S_1$ results from integrating out one flavour (say $i=1$), which is equal to the effective action in the GNY theory. It leads to a saddle solution where $\phi$ is uniform, and can be taken to zero at criticality. We thus have the dual of $N$ massless Dirac fermions with scaling dimension $\Delta_\psi=1$. It also follows that $\phi$ has dimension 1. 

\section{Offspring dualities}
We can apply formal transformations by gauging certain symmetries to obtain offspring dualities. In certain cases, we shall fall back on dualities that were
proposed using different approaches. 

\subsection{Single Dirac fermion}
We gauge the U$(1)_A$ symmetry, a process called $S$-duality, in the single flavour GNY transition \eqref{gn} (setting $h=1$):
\begin{align} \label{gn-gauge1}
    L_f' &= \bar\psi iD_b\psi - \phi\bar\psi\psi -r \phi^2 - \phi^4 +\frac{1}{2\pi}Bdb -\frac{1}{4\pi}BdB \\
    L_b' &=  |D_a z|^2 - |z|^4 + \frac{1}{4\pi}ada +\frac{1}{2\pi}b da  - \phi |z|^2 -r \phi^2 - \phi^4 +\frac{1}{2\pi}Bdb -\frac{1}{4\pi}BdB
\end{align}
More precisely, $B$ is a regular background gauge field, while $b$ is a dynamical spin$_c$ connection. We added $\frac{1}{2\pi}Bdb -\frac{1}{4\pi}BdB$ on both sides.
In the bosonic theory, $b$ appears linearly so we can integrate it out, which gives the constraint $a+B=0$, that allows to eliminate $a$:
\begin{align} \label{tri}
    L_b' &=  |D_{-B} z|^2 - |z|^4 - \phi |z|^2 -r \phi^2 - \phi^4 
\end{align}
The gauge field on the boson side has been eliminated, and we are left with a relatively simple theory of a complex boson coupled to a real scalar $\phi$. In fact, sitting at the critical point $r=0$, we could do as \cite{karch-more} and eliminate the interaction terms that appear above, see the discussion in Section~\ref{sec:1dirac}, to get
\begin{align} \label{tri2}
    L_b' &=  |D_{-B} z|^2 +\cdots 
\end{align}
where the dots denote potential terms for $z$ that are higher than quartic order. 
The duality involves a single-flavour gauged-GN transition in \eqref{gn-gauge1}, and a ``tricritical'' type theory of a regular complex scalar, written as \eqref{tri} or \eqref{tri2}. 

\emph{Phase diagram}---In order to discuss the phase diagram, the Yukawa form \eqref{tri} will be more useful to us. When $r>0$, $\phi$ is gapped, and we fall back on the duality that fermion QED with a single flavour is dual to a regular Wilson-Fisher boson~\cite{Seiberg2016}. For $r<0$, $\phi$ condenses giving the fermion theory \eqref{gn-gauge1} a mass term of either sign. When the condensate $\phi_0>0$, one gets a fully gapped theory due to the CS term at level 
$-1$ for $b$, which in turn generates a background term $\tfrac{1}{4\pi}BdB$. The background term already present in \eqref{gn-gauge1} cancels it exactly so that there is no net CS term for $B$ in the response. 
For the other sign $\phi_0<0$, integrating out $\psi$ gives no CS term, and we get a monopole condensate that spontaneously breaks the U(1) symmetry; the Goldstone mode is  realized by the photon, which has an implicit Maxwell term. To get the IR effective Lagrangian for $B$, we use the fact that $\tfrac{1}{2\pi}Bdb$ effectively sets $B=0$ so that no CS term for $B$ remains. 
On the boson side $\phi_0>0$ gives a mass gap, and the effective CS Lagrangian for $B$ vanishes, matching the fermion side. For $\phi_0<0$, the negative mass for $z$ leads to a spontaneous breaking of U(1) symmetry $\langle z\rangle\neq 0$ and a Goldstone mode, again in agreement with $L_f'$.  

It should be noted that the putative transition could be 1st order.

\subsection{Two Dirac fermions}
\subsubsection{Gross-Neveu and deconfined quantum critical point dualities}
Let us look at the $N=2$ GNY duality \eqref{gn2} in a slightly different formulation by decoupling the quartic term in the boson seed  Lagrangian $L_b$: $|z|^4\to\sigma |z|^2$:
\begin{align} \label{gn2-hs}
    L_f' &= L_f[\psi_1] + L_f[\psi_2] - \phi\bar\Psi\Psi  \\
    L_b' &= \hat L_b[\hat z_1,\hat a_1] + L_b[z_2,a_2] - \phi (-\hat\sigma_1 + \sigma_2)   \nonumber
\end{align}
where the time-reversed boson Lagrangian was used for flavour $1$:
\begin{align}
    \hat L_b[\hat z_1,\hat a_1] &=|D_{\hat a_1}\hat z_1|^2 -\hat\sigma_1 |\hat z_1|^2 -\frac{1}{4\pi}\hat a_1d\hat a_1-\frac{1}{2\pi}A_1 d\hat a_1 - \frac{1}{4\pi}A_1dA_1 \\
     L_b[z_2, a_2] &=|D_{a_2} z_2|^2 - \sigma_2 |z_2|^2 + \frac{1}{4\pi}a_2 da_2 + \frac{1}{2\pi}A_2 da_2  
\end{align}
where in this section we shall leave background gravitational CS terms implicit.  
Note that we have absorbed the coupling $h$ in \eqref{gn2-hs}. 
Since $\sigma$ is odd under timer-reversal, $\sigma\to -\hat\sigma$, we get $\bar\psi_1\psi_1 \leftrightarrow -\hat\sigma_1$ and $\bar\psi_2\psi_2\leftrightarrow \sigma_2$, which explains the interaction term in \eqref{gn2-hs}. Now, we can combine the potential terms on the boson side into
\begin{align}
    V_b= \hat\sigma_1|\hat z_1|^2 +\sigma_2|z_2|^2 + \phi(-\hat\sigma_1+\sigma_2) \to \sigma_2(|\hat z_1|^2+ |z_2|^2)+ \phi(-\hat\sigma_1+\sigma_2)
\end{align}
where we shifted $\phi\to \phi+|\hat z_1|^2$. We can then integrate out $\phi$, which sets $\hat\sigma_1=\sigma_2\equiv\sigma$, and the potential becomes
\begin{align}
    V_b = \sigma (|\hat z_1|^2+ |z_2|^2)
\end{align}
This looks like what we had in \eqref{gn2}, but with the difference that the quartic terms in $\hat L_b[1]$ and $L_b[2]$ have been eliminated. Explicitly:
\begin{align} 
    L_f' &= L_f[\psi_1;A_1] + L_f[\psi_2;A_2] - \phi(\bar\psi_1\psi_1+\bar\psi_2\psi_2)  \\
    L_b' &= |D_{\hat a_1}\hat z_1|^2  -\frac{1}{4\pi}\hat a_1d\hat a_1-\frac{1}{2\pi}A_1 d\hat a_1 - \frac{1}{4\pi}A_1dA_1+
    |D_{a_2} z_2|^2 + \frac{1}{4\pi}a_2da_2 +\frac{1}{2\pi}A_2 da_2 - \sigma (|\hat z_1|^2+|z_2|^2)   \nonumber
\end{align}

We can now perform a sequence of formal transformations on the terms involving gauge fields. Before doing so, we set $A\equiv A_1=A_2$. As a first step, we send $\tfrac{1}{4\pi}AdA$ to the fermion side. Second, on both sides, we $S$ dualize $A\to a$ and add $\tfrac{1}{2\pi}a dB+\tfrac{1}{4\pi}BdB$. \blu{This results in
\begin{align} \label{new-duality}
L_f' &= L_f[\psi_1;a] + L_f[\psi_2;a] - \phi \bar\Psi\Psi + \frac{1}{4\pi}ada +\frac{1}{2\pi}adB +\frac{1}{4\pi} BdB \\
     L_b' &=   |D_{a_2+B}\hat z_1|^2 + |D_{a_2} z_2|^2  -\sigma(|\hat z_1|^2+|z_2|^2) -\frac{1}{2\pi} a_2dB   \nonumber
\end{align}
where on the boson side we have eliminated both $a$ and $\hat a_1$ by using the fact that $a$ appeared linearly in the Lagrangian.
We have thus arrived at a new duality \eqref{new-duality} between fermion and boson QED poised at their respective critical points. Both theories have a single dynamical gauge field, in contrast to the two that we started with in $L_b'$. 
Had we set $\phi=0$ in our starting point \eqref{gn2-hs}, we would have obtained the duality between fermion QED (the phase, not the critical point), and boson easy-plane QED~\cite{review} with the two background gauge fields set to be equal. In our boson Lagrangian $L_b'$ \eqref{new-duality}, we recognize the manifestly SU(2)-invariant CP$^1$ model that describes the deconfined quantum critical point between N\'eel and VBS phases~\cite{dqcp,review}. Before analyzing the physical consequences, let us now see whether we can simplify the dual fermion theory $L_f'$. 
}

\blu{It turns out that we can do a transformation on $L_f'$ only. In the partition function, we can keep the path integral over $\phi$ to be performed last. For each $\phi$ configuration,  we thus have fermion QED with a symmetric mass source $\phi$. We can now exploit another form~\cite{Wang2017} of the duality for fermion QED and easy-plane boson QED, where no CS term appears for $a$ on the fermion side so that time-reversal is manifest:
\begin{align}
    L_f' &= L_f[\psi_1;a] + L_f[\psi_2;a] - \phi \bar\Psi\Psi + \frac{1}{4\pi}ada +\frac{1}{2\pi}adB +\frac{1}{4\pi} BdB \label{f-qed1}\\
     &\leftrightarrow  L_f[\psi_1;a] + L_f[\psi_2;a] - \phi \bar\Psi\Psi -\frac{1}{2\pi}adB - \frac{1}{4\pi} BdB   \label{f-qed2}
\end{align}
In fact, this is tantamount to a self-duality under gauging $B\to c$ and adding $\tfrac{1}{2\pi}c dB$ in \eqref{f-qed1}. The gauge field $c$ can then be eliminated via its equation of motion to yield \eqref{f-qed2}.} Additional details are given in Appendix~\ref{ap:sdual}. 
We thus obtain a new form of the duality \eqref{new-duality}:
\begin{align}  \label{dqcp}
    L_f'&= \bar\Psi i\slashed D_a\Psi- \phi\bar\Psi\Psi -\frac{1}{2\pi}adB-\frac{1}{4\pi}BdB  \\
    L_b' &= |D_{b+B} z_1|^2+|D_{b} z_2|^2- \sigma Z^\dag Z -\frac{1}{2\pi} B db 
\end{align}
where we relabelled fields. 
We recognise the duality~\cite{Wang2017} for the deconfined quantum critical point that describes the putative continuous transition between N\'eel and VBS states with emergent SO(5) symmetry~\cite{dqcp}. Although non-trivial evidence for the duality has been put forth, such as approximate matching monopole scaling dimensions~\cite{Dupuis2021-anomalous}, ultimately this transition may be (weakly) first order~\cite{review}. \blu{This is not a problem for our initial GNY duality,
as the formal manipulations ($S$ duality) are not always guaranteed to lead to CFTs.} 

\subsubsection{Ising Gross-Neveu}
We now consider the chiral Ising potential of Section~\ref{sec:ign}, with $V_f=\phi(\bar\psi_1\psi_1 - \bar\psi_2\psi_2)$, the boson potential becomes
\begin{align}
    V_b= \hat\sigma_1|\hat z_1|^2 +\sigma_2|z_2|^2 + \phi(-\hat\sigma_1- \sigma_2) \to \hat\sigma_1(|\hat z_1|^2 - |z_2|^2)+ \phi(-\hat\sigma_1 -\sigma_2)
\end{align}
where we shifted $\phi\to \phi+|z_2|^2$. We can then integrate out $\phi$, which sets $\hat\sigma_1=-\sigma_2\equiv\sigma$, and the potential simplifies:
\begin{align}
    V_b = \sigma (|\hat z_1|^2 - |z_2|^2)
\end{align}
\blu{By following the same procedure as was described above, we get another offspring duality
\begin{align} \label{new-duality-ising}
L_f' &= L_f[\psi_1;a] + L_f[\psi_2;a] - \phi \bar\Psi\sigma_z\Psi + \frac{1}{4\pi}ada +\frac{1}{2\pi}adB +\frac{1}{4\pi} BdB \\
     L_b' &=   |D_{a_2+B}\hat z_1|^2 + |D_{a_2} z_2|^2  -\sigma(|\hat z_1|^2 -|z_2|^2) -\frac{1}{2\pi} a_2dB   \nonumber
\end{align}
or by transforming $L_f'$ as above:}
\begin{align}  \label{gauged-ising}
    L_f'&= \bar\Psi i\slashed D_a\Psi- \phi (\bar\psi_1\psi_1 - \bar\psi_2\psi_2) -\frac{1}{2\pi}Bda-\frac{1}{4\pi}BdB  \\
    L_b' &= |D_{b+B} z_1|^2+|D_{b} z_2|^2-  \sigma (|z_1|^2 - |z_2|^2) -\frac{1}{2\pi} b dB  
\end{align}
This duality between a gauged Ising GNY theory and bosonic gauge theory (called QED$^-$), has been proposed in \cite{Cenke2017,Benvenuti2018}.

\section{Monopoles}
We will now obtain estimates of scaling dimensions of operators of charge $Q$ in the GNY CFT with $N=1$ flavour from the dual gauge theory. First, it will prove useful to decouple the quartic interaction with a Hubbard-Stratonovich field 
\begin{align} \label{karch}
    L_b'=|D_a z|^2 -\sigma |z|^2 +\frac{1}{4\pi}ada -\phi\sigma+\cdots
\end{align}
where we have replaced $|z|^2$ by $\sigma$ in the last term, and absorbed a coupling. The ellipsis denote terms that will play no role at low energy in our analysis. Eq.~\eqref{karch} is as in \cite{karch-more}, but we shall not integrate out any fields. Next, we consider the following generalisation of the theory to $N$ complex scalars and general CS level 
\begin{align} \label{karch-n}
    L_b'=\sum_i^N\left(|D_a z_i|^2 -\sigma |z_i|^2\right) +\frac{k}{4\pi}ada -\phi\sigma+\cdots
\end{align}
Going to imaginary time, taking the large-$N$ limit such that $\kappa=k/N$ remains finite, and rescaling $\phi$ by $N$, yields a saddle equation for $\phi$: $\sigma=0$. This remains true when we place the theory  on $S^2\times\mathbb R$, with the sphere pierced by flux $2\pi Q$, $Q\in\mathbb Z$, as required by state operator correspondence for monopole operators. It proves useful to work on the thermal circle $S^1_\beta$ instead of $\mathbb R$, in which case the zero temperature limit of the free energy yields the scaling dimension, $\Delta_Q=\lim_{\beta\to\infty} F_Q$.  
We need to allow for a non-trivial line integral (holonomy) of the gauge field $a$ around the thermal circle~\cite{Chester17}: $\alpha=i\beta^{-1}\int_{S_\beta^1}a$, which is real. We find a non-trivial solution to the holonomy saddle equation: $\alpha= -\lambda_Q(0)-\beta^{-1}\ln(\xi/(\xi+1))$, where $\lambda_j(\bar\sigma)=\sqrt{(j+1/2)^2-(Q/2)^2+\bar\sigma}$
is the single-particle energy of bosons with mass-squared $\bar\sigma$ on the sphere. The parameter $\xi=Q |\kappa| /(Q+1)$ encodes the dependence on both $Q$ (taken to be non-negative) and $\kappa=k/N$. The holonomy thus has an expression as in regular QED-CS~\cite{Chester17} but with $\bar\sigma$ set to 0. We can then evaluate the free energy using zeta-function regularization to get the scaling dimension $\Delta_Q=N\Delta^{(0)}_Q=\lim_{\beta\to\infty} F_Q $:
\begin{align}
     \Delta_Q^{(0)} = Q |\kappa| \sqrt{Q/2+1/4} + I_Q
\end{align}
where the square root is $\lambda_Q(0)$. $I_Q$ is a $\kappa$-independent series given in Appendix~\ref{ap:m}; interestingly, $I_Q$ also gives the scaling dimension of a defect monopoles in the theory of a free complex scalar~\cite{Pufu13}. 
We show the numerical values for the first 5 monopoles in the first column of Table~\ref{tab}. Setting $N=1$, we observe that the resulting scaling dimensions are reasonable: $\Delta_{Q>1}$ respect unitarity bounds, with $Q=1$ showing a very minor violation. They also obey strict convexity for all $Q$: $\Delta_{Q+Q'}>\Delta_Q+\Delta_{Q'}$, consistent with the conjecture that the charge spectrum of most CFTs will obey some form of convexity~\cite{convex}. 
The middle column corresponds to regular QED-CS where one tunes $|z|^2$ to reach criticality~\cite{Chester17,Chester21,Chester22}, while the last one is for single free Dirac fermion. Now, we want to compare the first column to the results for the $N=1$ GNY CFT.
It is always dangerous to extrapolate to $N=1$, but monopole large-$N$ results have proven surprisingly robust~\cite{Dupuis2021-anomalous,Chester22}. 
The rigorous conformal bootstrap result~\cite{archi} is
\begin{align}
    \Delta_\psi = 1.06861(12)
\end{align}
which compares reasonably well to our dual estimate $\Delta_1\approx 0.96$ in Table~\ref{tab}. A similar phenomenon occurs for the seed duality, where the leading order result (middle column) is actually equal to the exact answer (last column). It would be interesting to adapt the QED-CS calculation~\cite{Chester21,Chester22} for the quantum $\cO(N^0)$ correction to the present case. In addition, it would be interesting to obtain large-$N$ estimates for the other GNY theories discussed in this paper. 
\begin{table}[t]
\centering
\begin{tabular}{c||c||c||c}
$Q$ & $ \Delta^{(0)} $ &  $ \Delta_{\rm qed-cs}^{(0)}$ & $\Delta^{\rm free}$   \\ 
\hline
\hline
$1$ & 0.963 & 1 & 1  \\
$2$ & 2.46 &  2.58 & 2  \\
$3$ & 4.35 & 4.59 & 4    \\
$4$ & 6.57  & 6.94 & 6  \\
$5$ & 9.83  & 9.59 & 8  \\
\hline
\end{tabular}
\caption{Our results for the scaling dimensions $\Delta_{Q}=N\Delta_{Q}^{(0)}+O(N^0)$ for charge $Q$ monopole operators in the
bosonic QED Yukawa theory \eqref{karch-n} with $N$ complex scalars coupled to a gauge field at Chern-Simons level $k$, with $k/N=1$ in a large $N,k$ expansion. 
The middle column is the leading order
estimate for regular bosonic QED-CS \cite{Chester17,Chester21,Chester22}. The last column gives the  dimension of a charge $Q$ operator in the theory of a single free Dirac fermion.} 
\label{tab}
\end{table} 

\section{Conclusion and outlook}
We have conjectured dualities for quantum critical points of Dirac fermions by exploiting the seed duality \eqref{seed}. 
We have employed a UV complete Yukawa formulation on both sides, which allowed a rather transparent matching of symmetries, and phase diagrams.
By gauging certain symmetries, a generalisation of the particle-vortex transformation, we have arrived at dualities between gauge theories. In particular, our duality for the maximally symmetric O(4) GNY critical point let to the duality between boson QED (deconfined quantum critical point) and a fermion QED-GNY theory~\cite{Wang2017}. Finally, we performed a large-$N$ estimate for monopole scaling dimensions on the boson side, that are dual to powers of the fermion in the single-Dirac fermion GNY transition. 

As these are conjectures, it would be important to perform further checks. Many of the multi-component CS theories that we put forward have never been studied before, so it would be of interest to determine the scaling dimensions of the basic operators, including monopoles. Furthermore, it might be possible to simplify certain of these theories. It would be interesting to see if a connection can be made to the parallel proposal of the duality for the HGN theory \eqref{hgn}, where a single dynamical gauge field appears~\cite{xu23}. A similar construction appeared previously for the free CFT with 2 Dirac fermions~\cite{Wang2017}. Some properties are less manifest with these proposals, including the action of time-reversal, but they also have merits.  

Finally, it would be interesting to study other transitions, such as quantum critical points involving Majorana fermions. Starting with 
the conjectured duality between a single Majorana fermion and 3-component boson coupled to an SO(3) gauge field with CS level 1, the idea of \cite{karch-more} was applied to get the GNY transition with $N=1/2$ Dirac fermions, i.e.\ one 2-component Majorana fermion~\cite{Metlitski2017}. Using the Yukawa formulation on the boson side, one could construct other transitions as we did. In particular, for an even number of Majorana fermions, one could obtain alternate dualities for the transitions discussed in this work.     

\section*{Acknowledgments}
We thank Tarun Grover for sharing his unpublished results with us, for numerous discussions, and for providing valuable feedback. 
We also thank Shai Chester for countless discussions and critical feedback throughout the project. We are grateful to \'Eric Dupuis and AmirHossein Fallah Zarrinkar for  discussions, and collaboration on closely related topics. 
Finally, we thank Joseph Maciejko, Maxim Metlitski and Manu Paranjape for useful discussions and feedback.  WWK is supported by a
a grant from the Fondation Courtois, a Discovery Grant from NSERC, a Canada Research Chair, and Team Research Project from FRQNT. 

\appendix 
\section{Gauging dualities} \label{ap:sdual}
We describe the $S$ and $T$ moves on the $N=2$
GNY duality. \blu{We keep the gravitational CS terms implicit.} 
In step 1, we send $\tfrac{1}{4\pi}AdA$ to the fermion side,
\begin{align} 
    L_f' &= L_f[\psi_1;A] + L_f[\psi_2;A] - V_f + \frac{1}{4\pi}AdA \\
    L_b' &= |D_{\hat a_1}\hat z_1|^2 + |D_{a_2} z_2|^2  -V_b -\frac{1}{4\pi}\hat a_1d\hat a_1-\frac{1}{2\pi}A d\hat a_1 
    + \frac{1}{4\pi}a_2da_2 +\frac{1}{2\pi}A da_2    \nonumber
\end{align}
where we keep the potential terms general for now. 

2) On both sides, we $S$ dualize $A\to a$ and add $\tfrac{1}{2\pi}a dB+\tfrac{1}{4\pi}BdB$: 
\begin{align} 
    L_f' &= L_f[\psi_1;a] + L_f[\psi_2;a] - V_f + \frac{1}{4\pi}ada +\frac{1}{2\pi}adB +\frac{1}{4\pi} BdB \label{qed-f-1} \\
    L_b' &= |D_{\hat a_1}\hat z_1|^2 + |D_{a_2} z_2|^2  -V_b -\frac{1}{4\pi}\hat a_1d\hat a_1-\frac{1}{2\pi}a d\hat a_1 
    + \frac{1}{4\pi}a_2da_2 +\frac{1}{2\pi}a da_2  +\frac{1}{2\pi}adB  +\frac{1}{4\pi}BdB \nonumber
\end{align}
The dynamical gauge field $a$ is a spin$_c$ connection, while $B$ is a regular U(1) background gauge field.
We see that $a$ appears linearly in $L_b'$, implementing the constraint $\hat a_1=a_2+B$, which leads to 
\begin{align}
     L_b' = |D_{a_2+B}\hat z_1|^2 + |D_{a_2} z_2|^2  -V_b -\frac{1}{2\pi} a_2dB 
\end{align}

3) On the fermion side, \blu{we can exploit a self-duality of fermion QED whereupon the CS term for the dynamical gauge field is eliminated
\begin{align}
    & L_f[\psi_1;a] + L_f[\psi_2;a] + a^2+2aB+B^2 \label{dual1} \\
     \leftrightarrow \; & |D_{b} z_1|^2 + |D_{b-B} z_2|^2  -|z_1|^4 - |z_2|^4 +2bB \label{dual2} \\
    \leftrightarrow \; & L_f[\psi_1;a] + L_f[\psi_2;a]  -2aB - B^2 \label{dual3}
\end{align}
where we used a convenient polynomial notation for the gauge part
\begin{align}
    AB\equiv \frac{1}{4\pi} AdB 
\end{align}
Note that $AB=BA$ due to integration by parts.
The first duality for fermion QED \eqref{dual1} contains a CS term for $a$~\cite{review}, while the second one \eqref{dual3}
does not~\cite{Wang2017}. Eq.~\eqref{dual2}
corresponds to easy-plane bosonic QED. 
}

\blu{Interestingly, the above is equivalent to a self-duality of $L_f'$ in \eqref{qed-f-1} upon $S$ dualizing $B\to c$ and adding $\tfrac{1}{2\pi}c dB$:}
\begin{align} \label{almost}
     L_f' = L_f[\psi_1;a] + L_f[\psi_2;a] - V_f + a^2 +2ac + c^2 + 2cB  
\end{align}
where we have now reinstated the potential. \blu{We note that $c$ is a spin$_c$ connection.} 
The equation of motion for $c$ yields the relation $a+c+B=0$, allowing us to eliminate $c$. 
The gauge part in \eqref{almost} reads
\begin{align}
    a^2+2ac+c^2+2cB &\to a^2+2a(-a-B)+ (-a-B)^2 +2(-a-B)B\\
    &= -2aB+2aB+B^2-2aB-2B^2=-2aB-B^2
\end{align}
Putting everything together, we thus get the new duality
\begin{align}  \label{new}
    L_f'&= \bar\Psi i\slashed D_a\Psi- V_f -\frac{1}{2\pi}Bda-\frac{1}{4\pi}BdB  \\
    L_b' &= |D_{b+B} z_1|^2+|D_{b} z_2|^2- V_b -\frac{1}{2\pi} B db 
\end{align}
where we relabelled $\hat z_1\to z_1$ and $a_2\to b$.

Now, if we set 
\begin{align}
    V_f=0\,, \quad V_b = |z_1|^4 + |z_2|^4
\end{align}
we get the duality between easy-plane boson QED and fermion QED~\cite{Wang2017}, see above. 
This choice of potentials corresponds to starting with 2 free fermions, and using the seed duality, with the time-reversed version for one flavour. 

If instead we use
\begin{align}
    V_f=\phi(\bar\psi_1\psi_1+\bar\psi_2\psi_2)\,,\quad V_b=|z_1|^4 + |z_2|^4 + \phi (|z_1|^2+|z_2|^2)
\end{align}
we arrive at the DQCP duality in the main text \eqref{dqcp}. This choice of potentials corresponds to starting with the 2-flavour GNY duality~\eqref{gn2}.

\section{Monopole scaling dimensions} \label{ap:m}
As given in \cite{Pufu13}, the zeta-regularized series takes the form ($Q\geq 0$):
\begin{align}
    I_Q = 2\sum_{j=Q/2}^\infty \left[ (j+1/2)\sqrt{(j+1/2)^2+(Q/2)^2} -(j+1/2)^2 + (Q/2)^2  \right]
\end{align}
which vanishes when $Q=0$, as expected. 

\bibliographystyle{ssg}
\bibliography{sat}

\end{document}